\documentclass[10pt,aps,prb,showpacs,superscriptaddress,floatfix,
twocolumn]{revtex4}
\usepackage{amsmath}
\usepackage{citesort}
\usepackage{graphicx}
\usepackage{dcolumn}
\begin{document}
\title{Optical sum in Nearly Antiferromagnetic Fermi Liquid Model}
\author{E. Schachinger}
\email{schachinger@itp.tu-graz.ac.at}
\homepage{www.itp.tu-graz.ac.at/~ewald}
\affiliation{Institute of Theoretical and Computational Physics\\
Graz University of Technology, A-8010 Graz, Austria}
\author{J.P. Carbotte}
\affiliation{Department of Physics and Astronomy, McMaster University,\\
Hamilton, Ontario, L8S 4M1 Canada}
\date{\today}
\begin{abstract}
We calculate the optical sum (OS) and the kinetic energy (KE) for a
tight binding band in the Nearly Antiferromagnetic Fermi Liquid
(NAFFL) model which has had some success in describing the electronic
structure of the high $T_c$ cuprates. The interactions among electrons
due to the exchange
of spin fluctuations profoundly change the probability of occupation
$(n_{{\bf k},\sigma})$ of states of momentum {\bf k} and spin $\sigma$
which is the central quantity in the calculations of OS and KE. Normal and
superconducting states are considered as a function of temperature.
Both integrals are found to depend importantly on interactions
and an independent electron model is inadequate.
\end{abstract}
\pacs{74.20.Mn 74.25.Gz 74.72.-h}
\maketitle
\newpage
\section{Introduction}

The high $T_c$ oxides fall in the category of highly correlated systems.
A manifestation of this fact is that in the
underdoped regime there exists a pseudogap.\cite{r1,r2,r3,r4,r5}
Precisely how it is to be described
remains controversial.\cite{r5,r6,r7,r23a,r8,r9,r10,r11,r12,r13}
In certain theories it is closely related to
superconducting correlations,\cite{r5,r6,r7,r23a} and the
superconducting transition temperature $T_c$ is the temperature at which
phase coherence
is lost. In other theories the
pseudogap has its origin in completely different correlations
and is a manifestation of a
competing order such as a $d$-density wave (DDW).\cite{r8,r9,r10,r11,%
r12,r13} In any case interactions among charge carriers play an
important role in such systems and cannot be ignored in any realistic
approach to their properties even around optimum doping which is the
case of interest here.

While many aspects of the superconducting state can be understood
qualitatively on the basis of extensions of BCS theory,
particularly around optimum doping,
the search
for essential differences\cite{r14} has remained an important
avenue of investigation. In particular the idea of kinetic energy
as opposed to potential energy driven superconductivity
(i.e.: the kinetic energy is reduced at the transition temperature)
and its relation to the OS has
recently been given serious consideration in
theory\cite{r14,r15,r16,r17,r18,r19,r20,r21,r22,r23}
and experimentally.\cite{r24,r25,r26,r27,r28,r29}
An issue of importance is the relation of the
kinetic energy (KE) to the optical sum (OS). A recent paper\cite{r28}
has provided some insight into this relationship and has given a
comparison with experiment for the temperature variation of the OS
in the normal state, its change at $T_c$, and its further
evolution in the superconducting state. Correlations beyond
BCS pairing were not considered, however.
There have also been several recent studies of the
temperature dependence of the OS, experimentally for the system
LA$_{2-x}$Sr$_x$CuO$_4$, Ref.~\onlinecite{r42a}, and, theoretically,
using other models.\cite{r42b,r42c,r42d}

Within a BCS model, the condensation is potential energy driven and,
in fact, the KE increases because the probability of occupation
of the state {\bf k} $(n_{\bf k})$ becomes smeared at the Fermi
energy by the opening of the gap. This translates into a decrease in the
OS as compared with its normal state value. But this is opposite to
the behavior obtained experimentally in Ref.~\onlinecite{r26}.
However, as mentioned, the theoretical discussion of Ref.~\onlinecite{r28}
is based on a non interacting model in the normal state and includes
pairing correlations in the superconducting state only at the level
of BCS. More sophisticated formulations of the theory of
superconductivity could give different results.
Also, interactions can significantly modify the results even
in the normal state, as has been demonstrated recently by
Knigavko {\it et al.}\cite{r30} in a simplified model in which
the charge carriers are coupled to a single Einstein mode. While
only the normal state was considered, it was found that
boson hardening (softening) results in an increase (decrease)
in the OS and that interactions
play an important role in determining temperature dependences.

In this paper we study a tight binding model with an emphasis on the
effect interactions can have on the OS, particularly on its
temperature dependence and its relationship to the KE. Here,
the interactions among the charge carriers are treated in the
Nearly Antiferromagnetic Fermi Liquid (NAFFL) model.\cite{r31}
A review of its main properties and successes is given in
Ref.~\onlinecite{r31}. The model is phenomenological and falls into
the general class of boson exchange models where the interaction
between electrons proceeds through the exchange of spin
fluctuations.\cite{r32} The imaginary part of the spin susceptibility
replaces the phonon propagator of the classic electron-phonon
Eliashberg theory. The spin susceptibility could be calculated
from microscopic theory. More usually, however, it is fit to
experimental data, specifically to NMR in Ref.~\onlinecite{r32}.
The basic idea is that the doped metallic cuprates are near an
antiferromagnetic phase boundary and that coupling to spin
fluctuations is therefore strong.
While there is no consensus as to the validity of such a model when
applied to the oxides the NAFFL model has been widely discussed and
has had considerable
successes particularly in correlating
superconducting properties,\cite{r33,r34,r35,r35a,r35b,r35c} and has
been used to
give a detailed description of the optical properties\cite{r35,r36,%
r37,r38,r39,r40,r43a} in the high $T_c$ oxides at optimum
doping. In any case, the NAFFL model\cite{r41}
provides a convenient and specific
framework within which  we can study the effect of correlations on the
OS and on related properties.

In Sec.~\ref{sec:2} we summarize the basic equations that are needed to
compute the optical sum integral as well as the KE. They are a set
of three coupled generalized Eliashberg equations written for any
momentum {\bf k} in the two dimensional CuO$_2$ Brillouin zone.
They involve renormalized Matsubara frequencies, the renormalized
quasiparticle energies as well as the superconducting energy gap.
The interaction
between the charge carriers is mediated by the exchange of spin
fluctuations and involve the spin susceptibility. Fast Fourier
transforms (FFT) provide solutions which give a $d$-wave gap as
observed in experiments. In Sec.~\ref{sec:3} we apply our solutions to
evaluate the probability of occupation of the state of momentum
{\bf k} and spin $\sigma$ $(n_{{\bf k},\sigma})$ from which the OS
and the KE follow. Their temperature dependence is studied. Comparison
with the non interacting case is made and it is concluded that
interactions can profoundly modify results. Different behaviors
can result depending on the choice of microscopic parameters. In
Sec.~\ref{sec:4} we consider explicitly the superconducting state. Here,
again, generalized Eliashberg equations give variations with $T$
which are considerably different from earlier BCS  results.
When the electron-exchange boson interaction is taken as
temperature independent and independent of state, the OS is lower
in the superconducting than in the normal state. Nevertheless,
it can keep increasing with decreasing temperature in contrast
to BCS where it was found to decrease. This increase can be traced to the
underlying temperature dependence of the OS which depends on
details of the band structure and interactions involved, for
example on the model spin susceptibility. Further, if we take account
of the low energy gaping of the spin susceptibility
which is brought
about by the superconducting transition (a process which is not
operative in the normal state) the KE can be further decreased
and there is an additional increase in the OS which can effectively
increase faster than it does in the normal state. In this case
the KE in the superconducting state with low energy gaping of the
spin susceptibility can be less
than in the normal state without low energy gaping.
An important conclusion
of our work is that the observation of a faster increase in the OS with
decreasing temperature in the superconducting state than in the
normal state cannot unambiguously be taken
to be an indication of kinetic energy driven superconductivity
in contrast to a recent claim by van der Marel {\it et al.}\cite{r28}
In Sec.~\ref{sec:5} we provide a brief conclusion.

We use units in which $\hbar = c = 1$ throughout this paper.

\section{Formalism}
\label{sec:2}

\begin{figure}[tp]
\vspace*{-1cm}
  \includegraphics[width=9cm]{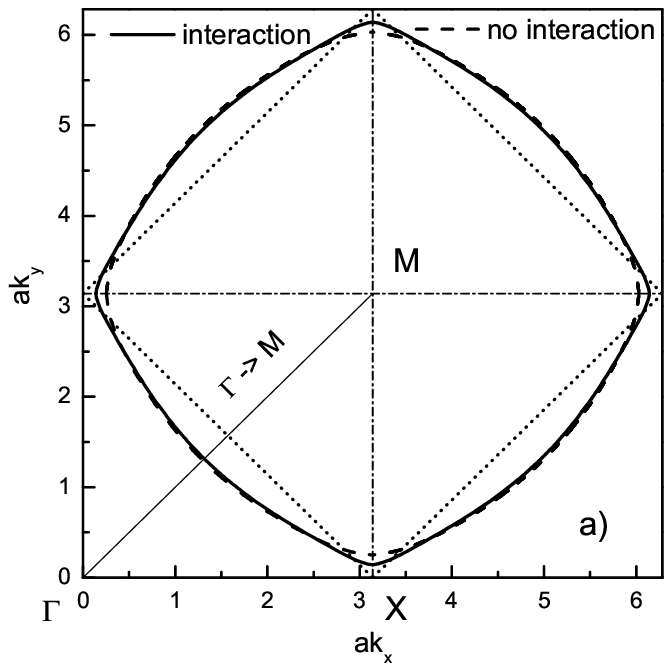}\\
\vspace*{-1cm}
  \includegraphics[width=9cm]{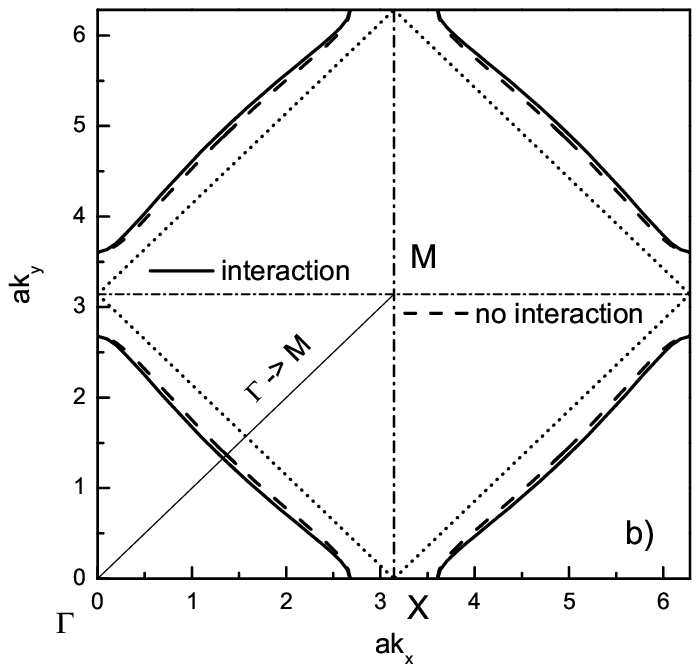}
\vspace*{-10mm}
  \caption{The Fermi surface for the non interacting system (dashed line)
and the system with interaction (solid line). a) model A of Table~\ref{tab:1},
b) for model B.}
  \label{fig:2}
\end{figure}

In the NAFFL model the interaction between holes proceeds through the
exchange of spin fluctuations and the spin susceptibility
$\chi({\bf q},\omega)$ plays a central role. The three Eliashberg
equations for the renormalized frequencies $\tilde{\omega}({\bf k},i\omega_n)$,
the energy renormalization $\xi({\bf k},i\omega_n)$ and the pairing energy
$\phi({\bf k},i\omega_n)$ as a function of momentum {\bf k} in the
two dimensional CuO$_2$ Brillouin zone, and of fermionic Matsubara
frequencies $i\omega_n = i\pi T(2n+1),\,n=0, \pm 1, \pm 2,\ldots$
and the temperature $T$ are\cite{r35a,r35b,r35c}
\begin{subequations}
\label{eq:1}
\begin{eqnarray}
  \tilde{\omega}({\bf k},i\omega_n) &=& \omega_n+
  T\sum\limits_m\sum\limits_{{\bf k}'}
  \lambda_{SF}({\bf k}-{\bf k}',i\omega_n-i\omega_m)¸\nonumber\\
  &&\times
  \frac{\tilde{\omega}({\bf k}',i\omega_m)}{D({\bf k}',i\omega_m)},
  \label{eq:1a}\\
  \xi({\bf k},i\omega_n) &=&
  -T\sum\limits_m\sum\limits_{{\bf k}'}
  \lambda_{SF}({\bf k}-{\bf k}',i\omega_n-i\omega_m)\nonumber\\
  &&\times
  \frac{\epsilon_{{\bf k}'}+\xi({\bf k}',i\omega_m)}{D({\bf k}',i\omega_m)},
  \label{eq:1b}\\
  \phi({\bf k},i\omega_n) &=&
  -T\sum\limits_m\sum\limits_{{\bf k}'}
  \lambda_{SF}({\bf k}-{\bf k}',i\omega_n-i\omega_m)\nonumber\\
  &&\times
  \frac{\phi({\bf k}',i\omega_m)}{D({\bf k}',i\omega_m)}.
  \label{eq:1c}
\end{eqnarray}
\end{subequations}
In Eqs.~\eqref{eq:1} $D({\bf k},i\omega_n)$ is given by
\begin{equation}
  \label{eq:2}
  D({\bf k},i\omega_n) = \tilde{\omega}^2({\bf k},i\omega_n)+
  \left[\epsilon_{\bf k}+\xi({\bf k},i\omega_n)\right]^2+
  \phi^2({\bf k},i\omega_n),
\end{equation}
and $\epsilon_{\bf k}$ is the charge carrier dispersion relation. In a
tight binding model without the inclusion of the coupling to the
spin fluctuations it is given by:
\begin{eqnarray}
  \epsilon_{\bf k} &=& -2\left\{t\left[\cos(ak_x)+\cos(ak_y)\right]
  \right.\nonumber\\ &&-\left.
  2t'\cos(ak_x)\cos(ak_y)\right\}-\mu^\ast.
  \label{eq:3}
\end{eqnarray}
Here, $a$ is the lattice parameter in the copper oxide plane,
$t$ the nearest neighbor hopping,
$t'$ the next nearest neighbor hopping, and $\mu^\ast$ the chemical
potential. We will discuss within this
context, two tight binding models with the parameters given in
Table~\ref{tab:1}. The corresponding Fermi surface is presented in
Fig.~\ref{fig:2}a for model A and in Fig.~\ref{fig:2}b for model B.
These figures define also the various points in the Brillouin zone.
The dotted lines indicate the antiferromagnetic Brillouin zone
boundary. For model A, the Fermi surface crosses the anti-ferromagnetic
\begin{table}[tp]
\caption{\label{tab:1}The two tight binding models used within this
paper. Model A corresponds to the tight binding model discussed by
van der Marel {\it et al.}\protect{\cite{r28}} $t$ and $t'$ are
given in meV, the critical temperature $T_c$ in K, and the filling
$\langle n\rangle$ is defined in Eq.~\protect{\eqref{eq:9}}.
}
\begin{ruledtabular}
\begin{tabular}{ldddd}
Model & t & t' & \langle n\rangle & T_c\\
\hline
A & 148.8 & 40.9 & 0.425 & 90\\
B & 100.0 & 16.0 & 0.4 & 100\\
\end{tabular}
\end{ruledtabular}
\end{table}
Brillouin zone around $X$ and symmetry related points. These points
are `hot spots' for which Fermi surface to Fermi surface transitions
are possible with momentum transfer $(\pi/a,\pi/a)$ (nesting
property). Model A has
been used previously in Ref.~\onlinecite{r28} to discuss the OS and
KE in Bi$_2$Sr$_2$Cu$_2$O$_{8+\delta}$ (BSCCO) assuming no interactions
between the electrons in the
normal state. Therefore, it is natural to employ this same model
in the present study which extends the previous calculations to
include exchange of spin fluctuations between charge carriers in
the NAFFL model. The second model, Model B, was chosen to contrast
with the first. It has no hot spots and its Fermi surface is closed
around the $\Gamma$-point in contrast to Model A which has a Fermi surface
which is closed around the $M$-point. It also has a smaller value of
$t$ which, on its own, would imply a smaller absolute value of KE.
These differences in band parameters lead, as we shall see, to
some differences in KE and optical sum at $T=0$ in the
normal state.

In the phenomenological model of Millis {\it et al.}\cite{r31,r32} and
Monthoux {\it et al.}\cite{r41} (MMP-model) the magnetic susceptibility
was fit to NMR data and its imaginary part takes on the form
\begin{equation}
  \label{eq:4}
  \Im{\rm m}\chi_{MMP}({\bf q},\omega) =
  \frac{\chi_{\bf Q}\,(\omega/\omega_{SF})}{\left[1+\zeta^2({\bf q}-{\bf Q})^2
  \right]^2+(\omega/\omega_{SF})^2},
\end{equation}
where $\chi_{\bf Q}$ is the static susceptibility, {\bf Q} is the
commensurate antiferromagnetic wave vector $(\pi/a,\pi/a)$
in the upper right hand quadrant of the CuO$_2$-plane Brillouin zone
and symmetry related points. $\zeta$
is the magnetic coherence length, and $\omega_{SF}$ a characteristic
spin fluctuation frequency. We set $\zeta = 2.5\,a$ throughout this
paper and various values for $\omega_{SF}$ are investigated.

The kernel $\lambda_{SF}({\bf q},i\nu_{n-m})$ in Eqs.~\eqref{eq:1}
with momentum transfer ${\bf q} = {\bf k}-{\bf k}'$ and the
bosonic Matsubara frequency $i\nu_{n-m} = i\omega_n-i\omega_m$ is
given as
\begin{equation}
  \label{eq:5}
  \lambda_{SF}({\bf q},i\nu_n) = \frac{g^2\chi_{\bf Q}}
  {1+\zeta^2({\bf q}-{\bf Q})^2+(\vert\nu_n\vert/\omega_{SF})},
\end{equation}
with $g^2\chi_{\bf Q}$ adjusted to get the desired value of the critical
temperature $T_c$ for a certain value of $\omega_{SF}$ from the
solution of the linearized Eqs.~\eqref{eq:1}. This defines the model.

The aim of this paper is to investigate the effect interactions have on
the OS defined as
\begin{equation}
  \label{eq:6}
   \pi\,e^2\,I_\sigma = 
  \int\limits_{-\Omega}^\Omega\!d\omega\,\Re{\rm e}\sigma_{xx}(\omega) =
  \frac{\pi\,e^2}{V}\sum\limits_{{\bf k},\sigma}
  n_{{\bf k},\sigma}\frac{\partial^2\epsilon_{\bf k}}{\partial k_x^2},
\end{equation}
where $e$ is the charge on the electron, $V$ the volume, and
$n_{{\bf k},\sigma}$ is the probability of occupation of a state of
momentum {\bf k} and spin $\sigma$. Finally, $\sigma_{xx}(\omega)$ is
the optical conductivity. The integral in Eq.~\eqref{eq:6} is to be
taken over the single band with $\Omega$, the upper limit in the
integral of Eq.~\eqref{eq:6}, large enough to include all
possible transitions in that band. We are also interested
in the relationship between the OS and the kinetic energy. By
definition
\begin{equation}
  \label{eq:7}
  I_{\rm KE} = \left\langle H_{\rm KE}\right\rangle = \frac{a^2}{V}
  \sum\limits_{{\bf k},\sigma} n_{{\bf k},\sigma}\epsilon_{\bf k}.
\end{equation}
We will see that, to a good approximation $I_\sigma$ and
$-I_{\rm KE}$ are nearly proportional to each other. Thus,
\begin{equation}
  \label{eq:8}
  \rho_L \propto \frac{1}{\pi e^2}\int\limits_{-\Omega}^\Omega\!
  d\omega\,\Re{\rm e}\,\sigma_{xx}(\omega) \approx -\frac{1}{2}\left\langle
  H_{\rm KE}\right\rangle.
\end{equation}
holds approximately. (An equal sign would be appropriate if the
dispersion relation \eqref{eq:3} contained only nearest neighbor
interaction, i.e.: $t'=0$.) Here, $\rho_L$ is the experimentally
determined value of the OS $(I_\sigma)$.

\begin{figure}[tp]
\vspace*{-3mm}
  \includegraphics[width=9cm]{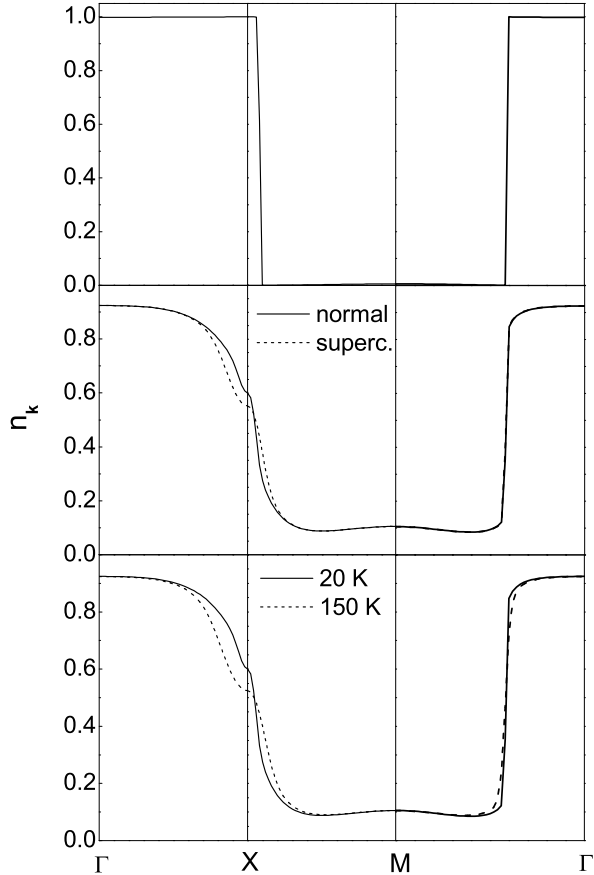}
  \caption{The occupation number $n_{\bf k}$ for both spin states for
selected directions in the CuO$_2$ Brillouin zone. Model A of
Table~\protect{\ref{tab:1}} was used.
Top frame: the non interacting case. Center frame: The interacting
case at a temperature $T=20\,$K. We show normal state (solid line)
and superconducting state (dashed line) results. Bottom frame: The
temperature influence on the normal state $n_{\bf k}$ for
$T=20\,$K (solid line) and $T=150\,$K (dashed line). }
  \label{fig:1}
\end{figure}

Interactions have a profound effect on the probability of occupation of
the state $\vert{\bf k},\sigma\rangle$ which would be one or zero for
occupied and unoccupied states respectively in the non interacting case.
In Fig.~\ref{fig:1} we show results for $n_{\bf k}$, along
certain selected
directions in the CuO$_2$ Brillouin zone. In all cases Model A of
Table~\ref{tab:1} is used for the
electronic dispersion \eqref{eq:3} with $\mu^\ast$ adjusted to the
required filling which is is defined as
\begin{equation}
  \label{eq:9}
  \langle n\rangle = \frac{1}{2}-\sum\limits_{\bf k}\sum\limits_{n\ge 0}
  \frac{\epsilon_{\bf k}+\xi({\bf k},i\omega_n)}
  {\tilde{\omega}^2({\bf k},i\omega_n)+\left[\epsilon_{\bf k}+
  \xi({\bf k},i\omega_n)\right]^2+\phi^2({\bf k},i\omega_n)},
\end{equation}
and the charge carrier spin fluctuation strength
$g^2\chi_{\bf Q}$ is adjusted to get a $T_c = 90\,$K for the
superconducting state. In the top frame of Fig.~\ref{fig:1} we show
$n_{\bf k}$ in the non interacting case as we go from $\Gamma$ to $X$
and from $X$ to $M$ in the Brillouin zone with the Fermi surface defined
as the value of {\bf k} at which $n_{\bf k}$ jumps from one to zero.
It is obvious from Fig.~\ref{fig:2}a (dashed line) that
the path $\overline{XM}$ crosses the Fermi surface. A second
crossing of the Fermi surface can be observed along the path from $M$
to $\Gamma$.

The center frame of Fig.~\ref{fig:1} shows $n_{\bf k}$
when interactions are taken into account. (The corresponding Fermi surface
is shown as the solid line in Fig.~\ref{fig:2}a.) We see a drastic
difference in the value of $n_{\bf k}$ which is now of the order
0.9 at the $\Gamma$ point indicating that the effect of the
interaction is very significant even in the center of the Brillouin zone.
Also, $n_{\bf k}$ is of the order 0.1 outside the non interacting
Fermi surface where $n_{\bf k}\approx 0$ for the non interacting case.
The solid line applies to the normal state at $T=20\,$K and the dashed
line to the superconducting state at the same temperature. On comparing
these two cases we see that the transition to the superconducting state
depletes even further the non interacting sea in the sense that it further
reduces $n_{\bf k}$ at ${\bf k}$s below the Fermi surface and
correspondingly increases the probability of occupation of states right
above the Fermi surface beyond the effect of interactions in the
normal state. It is to
be noted, however, that the onset of superconductivity results in rather
modest changes in $n_{\bf k}$ as compared to the difference between
interactions and no interactions in the normal state.

The bottom frame of Fig.~\ref{fig:1} gives results in the normal state
but compares two temperatures, namely $T=20\,$K (solid line) and
$T=150\,$K (dashed line). Comparison with the middle frame shows that
increasing the temperature has roughly the same qualitative effect
on $n_{\bf k}$ as does the transition to the superconducting state. In
both cases, the KE given by Eq.~\eqref{eq:7} increases because the states
of lower $\epsilon_{\bf k}$ get depleted while states with higher
$\epsilon_{\bf k}$ are occupied with increasing probability. This will
also hold for the OS according to Eq.~\eqref{eq:6} which will now
depend on interactions and on temperature.

\section{Results for the optical sum in the normal state}
\label{sec:3}

\begin{figure}[tp]
\vspace*{-5mm}
  \includegraphics[width=9cm]{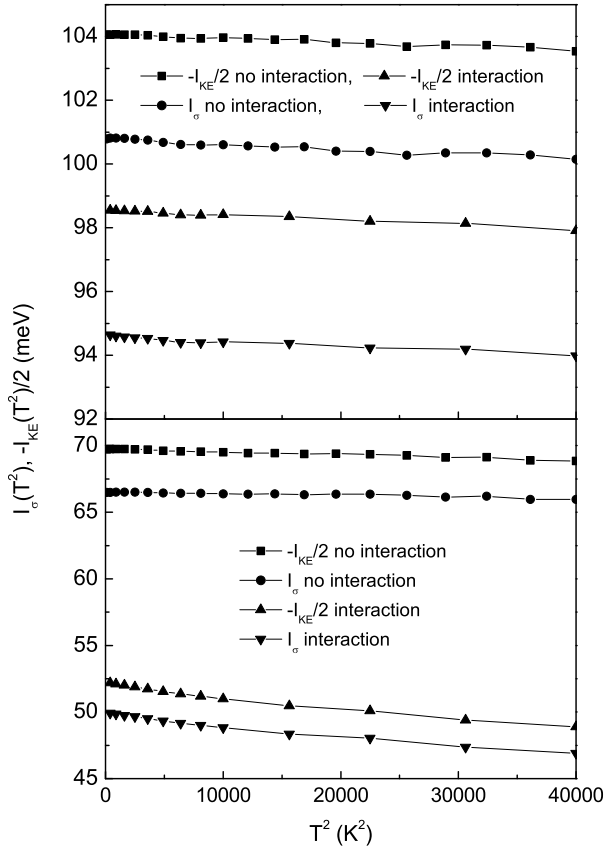}
  \caption{Optical sum $I_\sigma$ and kinetic energy, $-I_{\rm KE}/2$,
as a function of $T^2$. Solid circles and squares are for the non
interacting case while solid up-triangles and solid down-triangles
include interactions.
The top frame applies to
Model A of Table~\protect{\ref{tab:1}} and $\omega_{SF}=82\,$meV was
used. Here, the interacting and non interacting cases show similar
temperature dependencies. The bottom frame is for
Model B of Table~\protect{\ref{tab:1}} with an MMP-model
$\omega_{SF}=10\,$meV. Note the difference in temperature dependence
between interacting and non interacting case.
}
  \label{fig:4}
\end{figure}

In Fig.~\ref{fig:4} we show results for the
optical sum $(I_\sigma)$ Eq.~\eqref{eq:6} and compare with the kinetic
energy $(I_{\rm KE})$, Eq.~\eqref{eq:7}. The top frame is based on
Model A and the bottom frame on Model B of Table~\ref{tab:1}.
The solid squares and circles are $-I_{\rm KE}/2$
and $I_\sigma$ respectively in the free tight binding case, i.e.:
no interactions, plotted as a function of the square of the temperature
for the normal state. (A $128\times 128$ sampling of the
${\bf k}$-space, $ak_x, ak_y\in [0,\pi]$, was used but going to
a $256\times 256$ sampling did not influence the results.)
In both models variation with $T$ over the range $0$ to
$200\,$K is small (of order 1\% for Model A and 2\% for Model B)
as was also found in the work of
Molegraaf {\it et al.}\cite{r26} Also, the two integrals
$(I_\sigma$, $-I_{\rm KE}/2)$ track each
other closely even though they are not equal in magnitude.
(They would be equal for $t'=0$. We tried other Fermi surfaces, even
one with perfect nesting, i.e.: $t'=0$ and $\langle n\rangle = 0.5$,
and found no qualitative changes.) These results are for
comparison with results indicated by up/down-triangles
which include interactions in the NAFFL model.
The magnitude of both, the optical
sum integral and the kinetic energy has been changed considerably by
the interactions although the order remains the same, i.e.: $-I_{\rm KE}/2$
is greater than $I_\sigma$ and, again, they track each other.
More importantly, for the discussion here,
the temperature variation has been changed. Both integrals now
show variation of the order 8 to $9\%$. Also the dependence on
$T^2$ is not linear at small values of $T^2$. Model A shows similar
behavior for small values of $\omega_{SF}$.
It is clear that any estimate
based on the independent particle tight binding model is unreliable.
It is, however,
possible to chose specific parameters in the MMP-model which show
variations in the interacting case that are much closer to the non
interacting case. This is illustrated in the top frame of Fig.~\ref{fig:4}.
Here we used Model A of Table~\ref{tab:1}.
Again, results with and without interaction
are compared and both show little temperature variation. To get
this we used $\omega_{SF}=82\,$meV in our MMP form of Eq.~\eqref{eq:4}
without a change in the magnetic
coherence length. Both results, with and without interaction, show
little variation with temperature.
What this shows is that the magnitude as well as
the temperature variation of KE and of OS depends significantly on the
parameters used to characterize their electronic structure,
particularly the spin susceptibility.

We have done additional calculations for Model A with
$\omega_{SF} = 40,\,20,$ and $10\,$meV.
In all cases the change in KE due to interactions at $T=0$
increases with decreasing values of $\omega_{SF}$. In particular,
it changes at $T=0$ by 5.3\% when compared with the non interacting
case, for $\omega_{SF}=82\,$meV and by 15.8\% for $\omega_{SF}=10\,$meV.
The corresponding temperature changes from $T=0$ to $T=200\,$K are
roughly a factor of 5 smaller, more precisely, they are
0.7\% and 3.4\% respectively. 
Thus, a change in KE at $T=0$ due
to interactions also implies a corresponding change in temperature
dependence with both changes tracking each other.
For Model B the change in KE due to interactions is 34\% for
$\omega_{SF} = 10\,$meV with a 8.7\% increase in KE from $T=0$
to $T=200\,$K. These variations are about a factor of two larger
than for the equivalent case of Model A with a comparable value
of $\omega_{SF}$. Despite the fact that the two models
represent very different band structures $I_\sigma$ and
$-I_{KE}/2$ show essentially the same qualitative features in their
temperature and $\omega_{SF}$ dependence. However, the quantitative
differences are important.

\section{Results for the optical sum in the superconducting state}
\label{sec:4}

\begin{figure}[tp]
  \includegraphics[width=9cm]{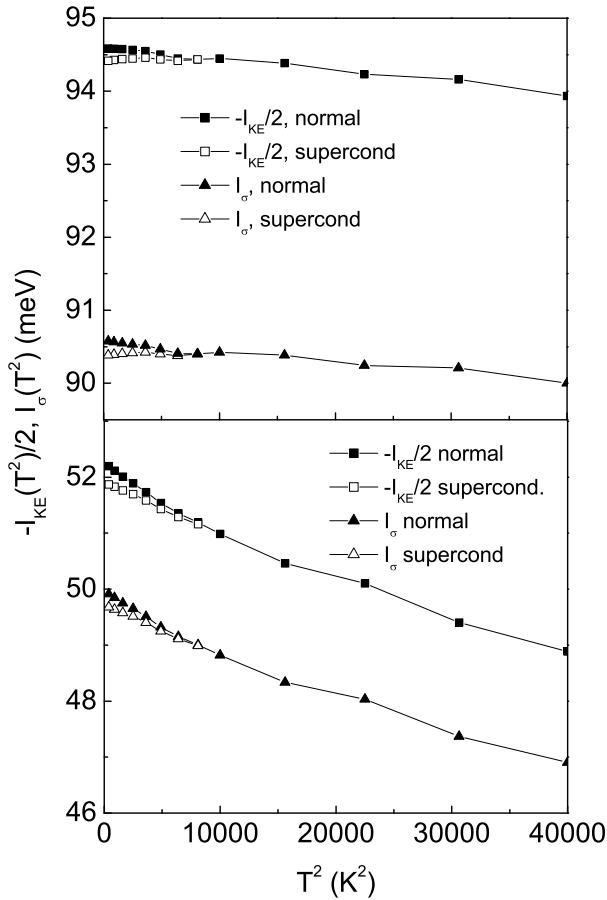}
  \caption{Comparison of normal and superconducting state for the
optical sum and the kinetic energy. The top frame applies to
Model A of Table~\ref{tab:1} and $\omega_{SF}=82\,$meV and the
bottom frame is for the band structure Model B of Table~\ref{tab:1} and
$\omega_{SF} = 10\,$meV.
}
  \label{fig:5}
\end{figure}

Results for the superconducting state are illustrated in Fig.~\ref{fig:5}
which has two frames. The top frame applies to the band structure
Model A of Table~\ref{tab:1} and is for $\omega_{SF}=82\,$meV
as in the top frame of Fig.~\ref{fig:4}. We have also included
2\% impurities in the
unitary limit but this serves mainly to illustrate that impurities
introduce no qualitative differences into our results. We see that,
as we expect, superconductivity reduces the optical integral
(open triangles) as compared with its normal state (solid
triangles) value at the same temperature.
This reduction is small. For the top frame which shows the least
temperature dependence, the KE integral shows a reduction of about
0.25\% below its normal state value which can be compared with the
results shown in the bottom frame of Fig.~6 of Ref.~\onlinecite{r29}
where the difference is 0.2\% in their BCS calculations. On the
other hand, in the bottom frame of our Fig.~\ref{fig:5} for Model
B the reduction is about 0.8\% (four times larger). This shows
that the Eliashberg results depend on band structure as well as on
the details of the interactions involved, in particular on the
value of $\omega_{SF}$. In the BCS limit the increase in KE
normalized to the absolute value of the condensation energy is
given by the formula $\left[\ln\left(\frac{\omega_D}{T_c}\right)%
-0.38\right]$ for both $s$- and $d$-wave superconductors. Here
$\omega_D$ is the Debye energy. This shows a strong dependence on
$\omega_D/T_c$. The formula itself, however, is valid only for
$\omega_D/T_c\gg 1$ and cannot be used to understand our Eliashberg
results. The NAFFL model includes interactions which,
as we have seen, change importantly the probability of occupation
$n_{\bf k}$ and consequently the optical integral as well as the
kinetic energy. For the parameters
of Model B and $\omega_{SF} = 10\,$meV we find
that $I_\sigma$ and $-I_{\rm KE}/2$ can keep increasing with
decreasing temperature in the superconducting state
(bottom frame of Fig.~\ref{fig:5}). The open squares and
triangles (superconducting state) are below their solid
counterparts (normal state) but still keep growing as the temperatures
is reduced. This does not indicate an exotic mechanism
but comes directly from a
generalization of Eliashberg theory that includes anisotropy in the band
structure and, more importantly, the interaction due to coupling to
the spin fluctuations.

\begin{figure}[tp]
  \includegraphics[width=9cm]{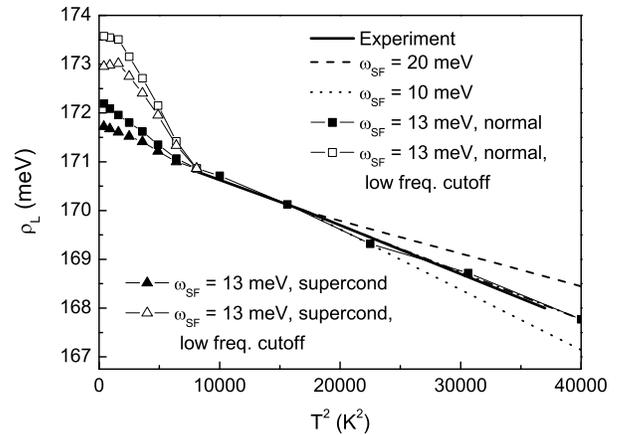}
  \caption{The optical sum as a function of the square of the temperature
for the band structure Model A of Table~\protect{\ref{tab:1}}
with different values of $\omega_{SF}$. Also in
one case a low frequency cutoff is applied to the spin susceptibility.
Note the significance of the $T^2$-variation on the value of
$\omega_{SF}$. The solid line indicates experimental normal
state results reported by Molegraaf {\it et al.}\protect{\cite{r26}}
}
  \label{fig:6}
\end{figure}

In Fig.~\ref{fig:6} we show additional results where the OS is seen to
increase even more rapidly, with reduced temperature below the onset
of superconductivity than it does in the normal state above $T_c$.
What is shown is $\rho_L$ which is $I_\sigma$ or $-I_{\rm KE}/2$
scaled to agree with experiment as discussed below; either,
$I_\sigma$ or $-I_{\rm KE}/2$, will do since these differ
mainly by a different scaling factor. Only
the OS is considered but the KE integral follows the same trend and
therefore the optical measurement can again be used to get information on
KE and its variation with temperature.

While in obtaining Fig.~\ref{fig:6} we applied Model A which
was used by van der Marel {\it et al.}\cite{r28} to describe
their optimally doped and underdoped BSCCO samples,
we now vary, in contrast to the top frame of Fig.~\ref{fig:5},
the value of $\omega_{SF}$ used in the MMP-model for
the spin susceptibility Eq.~\eqref{eq:4}. Results are presented
for $\omega_{SF}=20\,$meV (dashed line), $10\,$meV (dotted line),
$13\,$meV (solid squares) for the normal state, and
solid triangles for the
superconducting state. Also shown as the thick solid line are the
experimental results of Ref.~\onlinecite{r26} for their optimally
doped sample. We have scaled our
theoretical results to agree with experiment at $T=120\,$K. We first
note that varying $\omega_{SF}$ in the normal state can strongly
influence the temperature dependence obtained for $\rho_L$ (in meV).
 The value
of $\omega_{SF}=13\,$meV was chosen from a best fit in the region
$120\,{\rm K}\le T\le 200\,$K. The scaling factor required to get
agreement with this data is approximately 2. (When interactions are
neglected, as in Ref.~\onlinecite{r28}, the scaling factor is 
approximately 1.5.)
To reduce this
discrepancy, the value of $t$ would need to be increased but this
would also decrease the sensitivity of $I_\sigma(T)$ to temperature
variations and, thus, $\omega_{SF}$ would need to be adjusted as well.
To prove this we employ results of local density
approximation calculations by Markievicz {\it et al.}\cite{r42}
which suggest significantly bigger values for $t$ in BSCCO with
a Fermi surface which is little different from the one presented
in Fig.~\ref{fig:1}a. Using the dispersion relation, Eq.~(3),
and the parameter values of Table I of Ref.~\onlinecite{r42}
we find $I_\sigma(T=0) = 267.34\,$meV for the non-interacting system,
well above the value of $\rho_L(T=0) \approx 171.53\,$meV as has been
extrapolated from the normal state experimental data of
Ref.~\onlinecite{r26}. In order to reproduce the experimentally
observed temperature dependence $\rho_L(T)/\rho_L(T=0)$ we have
to introduce interactions and a value $\omega_{SF} = 8\,$meV is
found to give excellent agreement of $I_\sigma(T)/I_\sigma(T=0)$
with experiment. In this case $I_\sigma(T=0) = 230.34\,$meV, still
well above the experimental value and a down-scaling of
$I_\sigma(T)$ by a factor of 0.745 is required to achieve agreement
with experiment. Ultimately, a dispersion relation somewhere
between Model A and the one reported by Markievicz {\it et al.}\cite{r42}
and an $\omega_{SF}$ between $8$ and $13\,$meV will result in
a $I_\sigma(T)$ from theory which agrees with the experimental
$\rho_L(T)$ without scaling.
However, our main aim is not to treat a specific case but to
understand better the role interactions between the charge carriers
can play in the OS. Interactions introduce a new energy scale into
the problem, namely $\omega_{SF}$ for the NAFFL model. This
energy scale is additional to the chemical potential or the hopping
parameter $t$. With the values of the microscopic
parameters associated with the NAFFL model just described,
we proceed to compute
the OS for temperatures at and below $T_c$. The solid squares give
the continuation of the normal state curve and are presented for
comparison with the solid triangles which are the equivalent results
in the superconducting state. Again, superconducting state results
fall below the normal state ones but they are seen to, nevertheless,
increase with decreasing $T$. This occurs even if an Eliashberg
formulation is
used which represents a generalization of BCS theory and, in that
sense, is not exotic. The mechanism is the exchange of
antiferromagnetic spin fluctuations.
On comparing the top frame of Fig.~\ref{fig:5} with Fig.~\ref{fig:6},
we note that whether or not the superconducting state results keep increasing
with decreasing temperature is, for a given band structure, governed
by the value of $\omega_{SF}$.

In Fig.~\ref{fig:6} we show additional results, open squares for a
model normal state and open triangles for the superconducting
state. Now, there is a further dramatic increase in the OS both in the
normal and the superconducting state
as compared to its value in the normal state at $T=T_c=90\,$K. A
detailed explanation of how these results were arrived at is required.
In their analysis of optical data Carbotte {\it et al.}\cite{r34}
found that the spin fluctuations themselves are modified when the
superconducting state sets in. To carry out their analysis these
authors used a simplified version of our Eliashberg Eqs.~\eqref{eq:1}
which follows when the sum over {\bf k} is changed to an energy
integral as well as an angular average and the energy integral is done
analytically in a constant density of states approximation for an
infinite band model with interactions pinned to the Fermi surface.
Here we have been more realistic but what is important for us in
the work of Ref.~\onlinecite{r34} is that they find that the spin
fluctuation spectrum is gaped at low energies, or at
the very least loses intensity
and a spin resonance or peak forms at higher energy. This readjustment
in the spin susceptibility is not unexpected
and is a characteristic that should
be seen in any electronic mechanism for superconductivity.\cite{r37,%
r38,r39,r40,r43a,r41,r44,r45,r46,r47,r48}
Details are not important for the present discussion beyond the fact
that some adjustment of the spin susceptibility
$\Im{\rm m}\chi({\bf q},\omega)$ at small $\omega$ is expected,
which weakens the inelastic scattering.
Here we simply use the same low $\omega$ $[\omega_c(T)]$ cutoff 
applied to Eq.~\eqref{eq:4} which was determined by
E.~Schachinger {\it et al.},\cite{r43a} through consideration of
microwave data. Another approach would be to calculate the low
energy gaping of the spin susceptibility
from first principles but this would go beyond the scope of this
work and would introduce additional uncertainties.
The temperature dependence of $\omega_c(T)$
follows the temperature dependence of the superconducting gap
with a maximum value of $24\,$meV.
Application of this cutoff in
otherwise standard Eliashberg calculations based on our
Eqs.~\eqref{eq:1} yield the open triangles (superconducting state)
and open squares (normal state) of Fig.~\ref{fig:6}. The physics
underlying these curves has been made clear in a simple model
recently studied by Knigavko {\it et al.}\cite{r30} These authors
studied a model in which the charge carriers are coupled to a single
Einstein mode of unspecified origin. What they found was that stiffening
of this mode decreases the kinetic energy and hence increases the OS.
This is precisely the same mechanism that is operative in Fig.~\ref{fig:6}.
By applying a low frequency cutoff to the spin fluctuations in our
MMP-model we are decreasing the KE. This decrease in KE, present in the
underlying normal state below $T_c$, compensates for the increase in
KE intrinsic to the superconducting transition which results from the opening
of the superconducting gap. We note, however, that in our formulation
the OS, at any given temperature, is always below (although not very
much) its normal state value at this same temperature calculated with the
spectrum with a low frequency cutoff (open triangles). But this cutoff
is only operative in the superconducting state and is responsible for
making the open triangles fall above the solid squares.
The kinetic energy in the superconducting state with low
frequency cutoff is now less than the normal state kinetic energy
without cutoff.
Including the feedback mechanism of the formation of the
superconducting state on the spin susceptibility itself has the
net effect, at zero temperature (where it is largest),
of changing the sign of the KE
contribution to the condensation energy from that in BCS.

\section{Conclusion}
\label{sec:5}

We have used the Nearly Antiferromagnetic Fermi Liquid model to study
the effect of interactions on the optical sum and on the kinetic energy
in tight binding bands. Comparison of normal state results with
equivalent results when interactions are neglected showed that
temperature variations can be strongly affected by details of the
microscopic parameters involved in the spin fluctuation exchange
mechanism. Behaviors are possible which
can be quite different from the non interacting independent
particle model. Comparison with normal state
experimental data proves that the tight binding model of non
interacting particles is certainly not adequate to describe
properly the temperature dependence of the optical sum. (This
has also been observed by Benfatto {\it et al.}\cite{r42b}) Taking
into account interactions between the charge carriers makes the
tight binding model a viable model for the analysis of the
temperature dependence of the normal state optical sum. This
was demonstrated for the particular case of BSCCO and a particle
interaction modeled on the NAFFL. Other models,
like the one presented by Toschi {\it et al.}\cite{r42c} are also
capable to reproduce the temperature dependence
$I_\sigma(T)/I_\sigma(T=0)$ but they lack agreement with the
value of the optical sum at zero temperature.

When superconductivity is considered within an
Eliashberg formalism, the superconducting gap has $d$-wave
symmetry as a function of momentum in the two dimensional
Brillouin zone. The optical sum is found to decrease
with decreasing temperature for some range of parameters characterizing
the spin susceptibility but can also increase. This increase
cannot necessarily be interpreted as kinetic energy driven superconductivity.
In fact, in all cases considered, the optical sum is always lower,
at a given temperature in the superconducting state, than it is
in the corresponding normal state but, in some cases not by much.
Correspondingly, the kinetic energy is increased in the
superconducting state. What makes the optical sum and KE integral
continue to go up (in some cases) with decreasing temperature
is the fact that the interactions themselves introduce a
temperature dependence in the underlying normal state.

The results just described were obtained for a fixed
(i.e.: temperature independent) value
of the spin susceptibility. If
we consider the possibility that the spin fluctuation spectrum may itself
be modified\cite{r33,r34,r35,r44,r45,r46,r47,r48} by the onset of
superconductivity
and by temperature, even larger increases in the optical sum with
decreasing temperature can be obtained.
It is widely recognized that a generic feature of an electronic
mechanism of superconductivity is the possible gaping of the
excitation spectrum itself at small energies due to the opening
of the superconducting gap. This leads to a weakening of
interactions at small $\omega$ and to the so called collapse of
the inelastic scattering rate\cite{r44,r45,r46,r47,r48} which manifests
itself as a large peak in the temperature dependence of the
microwave conductivity. The weakening of the interaction
in the superconducting state through the opening of a low energy
gap in the spin susceptibility
corresponds to a decrease in KE in the superconducting state which can,
in the case considered, more than compensate for the intrinsic
increase that accompanies the formation of Cooper pairs
and, consequently, the OS rises with a larger slope in the
superconducting state than in the normal state just above $T_c$.
At $T_c$ there is no low energy gaping of the spin susceptibility and,
therefore, the mechanism for
KE reduction just described is not operative. In this sense
our model does not describe KE driven superconductivity.

\section*{Acknowledgment}
 
Research supported by the Natural Sciences and Engineering
Research Council of Canada (NSERC) and by the Canadian
Institute for Advanced Research (CIAR).

\end{document}